\documentclass[english,10pt]{article}

\usepackage{amsmath}
\usepackage{amssymb}
\usepackage{amsfonts}
\usepackage{amscd}
\usepackage{latexsym}
\usepackage{revsymb}
\usepackage{graphicx}

\newtheorem{defi}{Definition}

\newtheorem{thm}[defi]{Theorem}

\newtheorem{rem}[defi]{Remark}

\newtheorem{exempel}[defi]{Example}

\newenvironment{expl}[1][{}]{\begin{exempel} {#1}\normalfont}{\end{exempel}}
\newlength{\blank}
\newlength{\equalsign}
\newenvironment{beweis}[1][{\hspace{-\blank}}]{{\noindent\emph{Proof~{#1}.\ }}}{\hfill $\Box$\vskip 0.5\baselineskip}

\newcommand{\qed}{\hfill $\Box$}
\newcommand{\C}{{\mathbb{C}}}

\newcommand{\1}{{\openone}}

\newcommand{\ket}[1]{{|{#1}\rangle}}
\newcommand{\bra}[1]{{\langle{#1}|}}
\newcommand{\ketbra}[1]{{\ket{#1}\!\bra{#1}}}
\newcommand{\tr}{{\operatorname{Tr}}}
\newcommand{\rank}{{\operatorname{rank}\,}}

\begin{document}

\title{Remarks on additivity of\protect\\
       the Holevo channel capacity\protect\\
       and of the entanglement of formation}
\author{Keiji Matsumoto\thanks{ERATO Quantum Computation and Information project, Dai--ni Hongo White Bldg. 201, Hongo 5--28--3, Bunkyo--ku, Tokyo 133--0033, Japan. Email: {\tt keiji@qci.jst.go.jp}},\quad %
Toshiyuki Shimono\thanks{Department of Computer Science, Graduate School of Information Science and Technology, University of Tokyo, Hongo 7--3--1, Bunkyo--ku, Tokyo 113--0033, Japan. Email: {\tt shimono@is.s.u-tokyo.ac.jp}}\quad and\ \,%
Andreas Winter\thanks{Department of Computer Science, University of Bristol, Merchant Venturers Building, Woodland Road, Bristol BS8 1UB, United Kingdom. Email: {\tt winter@cs.bris.ac.uk}}}
\date{($16^{\rm th}$ May 2003)}
\maketitle


\begin{abstract}
  The purpose of this article is to discuss the relation between the additivity
  questions regarding the quantities \emph{(Holevo) capacity of a quantum channel $T$}
  and \emph{entanglement of formation of a bipartite state $\rho$}.
  In particular, using the Stinespring dilation theorem, we give a formula
  for the channel capacity involving entanglement of formation. This
  can be used to show that additivity of the latter for some states
  can be inferred from the additivity of capacity for certain channels.
  \par
  We demonstrate this connection for some families of channels,
  allowing us to calculate the entanglement cost for many states, including some
  where a strictly smaller upper bound on the distillable entanglement
  is known. 
  Group symmetry is used for more sophisticated analysis,
  giving formulas valid for a class of channels.
  This is presented in a general framework, extending recent findings of
  Vidal, D\"ur and Cirac.
  \par
  We also discuss the property of \emph{superadditivity}
  of the entanglement of formation, which would imply both the general additivity
  of this function under tensor products and of the Holevo capacity (with or without
  linear cost constraints).
\end{abstract}

\section{Introduction}
\label{sec:intro}
Quantum information theory has progressed considerably over the last decade:
today we understand much better the information transmission properties
of quantum channels, and entanglement has turned from an oddity first into
a valuable effect and then into a quantifiable resource,
as shown by the many well--motivated entanglement measures 
that have been put forward. Almost all of them are operationally grounded
as some optimal performance parameter, and can be written as solutions
to various high--dimensional or even asymptotic optimisation problems.
\par
All of these capacities and entanglement measures raise the natural problem
of \emph{additvity} under tensor products, i.e. the question, if the independent
supply of two specimens of the resource has as its performance the sum of the
performances of the individual objects (be they channels or states).
For some of the current measures of entanglement additivity has been
disproved by counterexamples (for the so--called \emph{relative entropy of
entanglement} in~\cite{VW}), for others, like the distillable
entanglement~\cite{BDSW} it is claimed improbable~\cite{evidence}.
For some, however, additivity is still widely conjectured, most notably
for a bound on the distillable entanglement by Rains~\cite{Rains}, and for
the \emph{entanglement of formation}~\cite{BDSW}.
\par
The literature on the subject is vast and increasing rapidly, and in the
present paper we will only make a small contribution. We shall be concerned
with the entanglement of formation, and with the
aforementioned classical capacity of quantum channels, pointing out
a connection between the two that also relates their additivity problems.
\par
We outline briefly the content of the rest of the paper: in
sections~\ref{sec:hcap} and~\ref{sec:entf} the classical capacity
of a channel and the entanglement of formation of a state are reviewed.
In section~\ref{sec:stine} a simple observation on the Stinespring
dilation of a completely positive map provides the link between the
two quantities, which is exploited in a number of examples
in section~\ref{sec:examples};
group symmetry is introduced in section~\ref{sec:symmetry},
adding another example, and to supply formulas valid for a class of channels
which includes examples discussed in section~\ref{sec:stine} as special cases.
And in section~\ref{sec:gap} some of these results are used to demonstrate
a gap between entanglement cost and distillable entanglement.
\par
In section~\ref{sec:superadd} we discuss \emph{superadditivity}
of entanglement of formation as a (conjectured) property which
would unify the additivity questions considered here: it implies
additivity of entanglement of formation, of channel capacity, and
of channel capacity with a linear cost constraint.
We conclude with a discussion of our observations and related works.

\section{Holevo capacity}
\label{sec:hcap}
We consider block coding of classical information via the quantum channel
$$T:{\cal B}({\cal H}) \longrightarrow {\cal B}({\cal H}_2),$$
where ${\cal H}$ and ${\cal H}_2$ are Hilbert spaces.
If the encoding is restricted to \emph{product states} it is
known~\cite{Holevo98,SW:97}
that the capacity is given by
\begin{equation}
  \label{eq:hcap}
  C(T)=\sup\left\{ I\bigl(p;T(\pi)\bigr) : \{p_i,\pi_i\}
                                    \text{ pure state ensemble on }{\cal H}\right\},
\end{equation}
where the \emph{Holevo mutual information} of an ensemble $\{p_i,\rho_i\}$
is given by
$$I(p;\rho)=S\left(\sum_i p_i\rho_i\right)-\sum_i p_i S(\rho_i).$$
Here $S(\omega)=-\tr\omega\log\omega$ is the von Neumann entropy of
a state. For finite dimensional ${\cal H}_2$
the $\sup$ in eq.~(\ref{eq:hcap}) is indeed a $\max$, attained for an ensemble
of at most $(\dim{\cal H}_2)^2$ states.
\par
It is conjectured that for a product of channels making use of entangled
input states does not help to increase the capacity:
\begin{equation}
  \label{eq:hcap:add}
  C(T_1\otimes T_2) = C(T_1)+C(T_2).
\end{equation}
(The question is implicit in~\cite{Holevo77} and the above references,
and made explicit in~\cite{BFS}, where it was speculated that the answer
may be negative.)
\par
This would imply that $C(T)$ is the classical capacity of $T$.
Observe that here the inequality ``$\geq$'' follows immediately
from the fact that the right hand side can be achived using product states.
Without additivity, the general formula for this capacity reads
$$\lim_{n\rightarrow\infty} \frac{1}{n}C\bigl(T^{\otimes n}\bigr).$$
Despite much recent activity on the question~\cite{AHW,AH},
and even proofs of the additivity conjecture in some
cases~\cite{bruss:etal,king:ruskai,fujiwara:hashizume,king:unital:add,%
king:unital:add:2,Shor02,king:depol}, it is still a wide open problem.

\section{Entanglement of formation}
\label{sec:entf}
Let $\rho$ be a state on ${\cal H}_1\otimes{\cal H}_2$. The
\emph{entanglement of formation} of $\rho$ is defined as
\begin{equation}
  \label{eq:entf}
  E_f(\rho):=\inf\left\{ \sum_i p_i E(\pi_i) : \{p_i,\pi_i\}
                    \text{ pure state ens.~with }\sum_i p_i\pi_i=\rho \right\},
\end{equation}
where the \emph{(entropy of) entanglement} for a pure state $\pi$ on
${\cal H}_1\otimes{\cal H}_2$ is defined as
$$E(\pi):= S\left(\tr_{{\cal H}_2}\pi\right)
         = S\left(\tr_{{\cal H}_1}\pi\right).$$
If the rank of $\rho$ is finite the $\inf$ is in fact a $\min$,
achieved for an ensemble of at most $(\rank\rho)^2$ elements.
\par
This quantity was proposed in~\cite{BDSW} as a measure of how costly
in terms of entanglement the creation of $\rho$ is.
\par
It is conjectured (but only in a few cases proved: the only published
examples are in~\cite{VDC}) that $E_f$ is an additive function with respect
to tensor products:
\begin{equation}
  \label{eq:entf:add}
  E_f(\rho_1\otimes\rho_2) = E_f(\rho_1)+E_f(\rho_2).
\end{equation}
Observe that, as in the case of the Holevo capacity, ``$\leq$''
follows easily from the fact that the right hand side is achieved by
product state ensembles. If this would turn out to be true, the
\emph{entanglement cost} $E_c(\rho)$ of $\rho$, i.e. the asymptotic rate of
EPR pairs to approximately create $n$ copies of $\rho$
is given by $E_f(\rho)$: in~\cite{HHT} it was proved rigorously that
$$E_c(\rho) = \lim_{n\rightarrow\infty} \frac{1}{n}E_f\bigl(\rho^{\otimes n}\bigr).$$
\par
Note that the function $E_f$ has the property of being a
\emph{convex roof}:
\begin{equation}
  \label{eq:entf:convroof}
  E_f(\rho) = \inf\left\{ \sum_i p_i E_f(\rho_i) : \{p_i,\rho_i\}
                    \text{ ensemble with }\sum_i p_i\rho_i=\rho \right\}.
\end{equation}
The cases in which $E_f$ is known are arbitrary states of
$2\times 2$--systems~\cite{wootters}, isotropic states in arbitrary
dimension~\cite{TV}, Werner and OO--symmetric states~\cite{VW},
and some other highly symmetric states~\cite{VDC}.

\section{Stinespring dilations: linking $C(T)$ and $E_f(\rho)$}
\label{sec:stine}
Due to a theorem of Stinespring~\cite{Stinespring} the
completely positive and trace preserving map $T$ can be presented as
the composition of an isometric embedding of ${\cal H}$ into a bipartite
system with a partial trace:
\begin{equation}
  \label{eq:embedding}
  T:{\cal B}({\cal H})
       \stackrel{U}{\hookrightarrow} {\cal B}({\cal H}_1\otimes{\cal H}_2)
                     \stackrel{\tr_{{\cal H}_1}}{\longrightarrow} {\cal B}({\cal H}_2).
\end{equation}
See~\cite{ruskai:review} for a discussion on how to construct this from
the so--called Kraus (operator sum) representation~\cite{kraus:book},
$T(\rho)=\sum_i A_i\rho A_i^*$ with $\sum_i A_i^* A_i=\1$, of $T$.
We shall use this construction later
on in the examples~\ref{expl:general:depol} and~\ref{expl:d-dim:depol}.
\par
By embedding into larger spaces we can present $U$ as restriction of
a unitary, which often we silently assume done.
Denote ${\cal K}:=U{\cal H}\subset{\cal H}_1\otimes{\cal H}_2$, the image
subspace of $U$. Then we can say that $T$ is equivalent to the partial
trace channel, with inputs restricted to states on ${\cal K}$.
This entails:
\begin{thm}
  \label{thm:hcap:entf}
  \begin{equation}
    \label{eq:hcap:alt}
    C(T) = \sup\{ S\bigl(\tr_{{\cal H}_1}\rho\bigr)-E_f(\rho) :
                                                     \rho\text{ state on }{\cal K}\}.
  \end{equation}
\end{thm}
\begin{beweis}
  Very simple: choosing an input ensemble for $T$ amounts by our
  above observation to choosing an ensemble $\{p_i,\pi_i\}$ on ${\cal K}$.
  Denoting $\rho=\sum_i p_i\pi_i$, the average output state of $T$ in
  eq.~(\ref{eq:hcap}) is just $\tr_{{\cal H}_1}\rho$, while the indivual
  output states are the $\tr_{{\cal H}_1}\pi_i$. Hence the second term
  in eq.~(\ref{eq:hcap}), the average of output entropies, has as its
  infimum $E_f(\rho)$ when we vary over ensembles with fixed $\rho$.
\end{beweis}
\par
Note that if we choose the dimension of ${\cal H}_1$ large enough,
every channel from ${\cal H}$ to ${\cal H}_2$ corresponds to a subspace
of ${\cal H}_1\otimes{\cal H}_2$ (though not uniquely) and vice versa.
\begin{rem}
  \label{rem:connes:etal}
  The quantity $S\bigl(\tr_{{\cal H}_1}\rho\bigr)-E_f(\rho)$ in the optimisation
  problem in theorem~\ref{thm:CC:implies:EE} equals the \emph{entropy
  of the subalgebra ${\cal B}({\cal H}_2)$ in ${\cal B}({\cal H}_1\otimes{\cal H}_2)$},
  as defined by Connes, Narnhofer and Thirring~\cite{CNT}: this was observed
  by Benatti, Narnhofer and Uhlmann~\cite{BNU}.
\end{rem}
\par\medskip
This has interesting consequences: for each subspace ${\cal K}$ of
the tensor product there is a convex set ${\cal O}_T$ of states $\rho$ supported
on it which maximise eq.~(\ref{eq:hcap:alt}). The reason for convexity
is again very simple: let $\rho,\rho'\in{\cal S}({\cal K})$. Then
\begin{align*}
  S\bigl( p\tr_{{\cal H}_1}\rho+(1-p)\tr_{{\cal H}_1}\rho' \bigr)
                            &\geq pS(\tr_{{\cal H}_1}\rho)+(1-p)S(\tr_{{\cal H}_1}\rho'), \\
  E_f\bigl( p\rho+(1-p)\rho' \bigr)
                            &\leq pE_f(\rho)+(1-p)E_f(\rho'),
\end{align*}
by concavity (convexity) of $S$ ($E_f$). Hence the aim function in
eq.~(\ref{eq:hcap:alt}) is concave, which implies that the set of $\rho$
for which it is at least $R$ is a convex set, for any real $R$.
\par
Observe that by this argument both $S\bigl(\tr_{{\cal H}_1}\rho\bigr)$
and $E_f(\rho)$ are constants for $\rho\in{\cal O}_T$. Indeed,
one can show (see the discussion below, in this section) that even
all $\tr_{{\cal H}_1}\rho$, $\rho\in{\cal O}_T$, are identical.
\par
For such states the additivity of $E_f$ is implied by the
additivity of $C$ for the corresponding channels:
indeed, assume that for two channels $T$, $T'$ that optimal
input states in the sense of eq.~(\ref{eq:hcap:alt})
are $\rho\in{\cal O}_T$, $\rho'\in{\cal O}_{T'}$,
respectively, with reduced states $\rho_2$ and $\rho_2'$.
Then, assuming additivity we get
\begin{equation}
  \label{eq:additivity:clue}
  \begin{split}
  S(\rho_2)-E_f(\rho) + S(\rho_2')-E_f(\rho')
               &=    C(T)+C(T')     \\
               &=    C(T\otimes T') \\
               &\geq S(\rho_2\otimes\rho_2')-E_f(\rho\otimes\rho'),
  \end{split}
\end{equation}
hence
$$E_f(\rho\otimes\rho') \geq E_f(\rho)+E_f(\rho'),$$
which by our earlier remarks implies additivity. Thus we have proved
\begin{thm}
  \label{thm:CC:implies:EE}
  If for any two channels $T$ and $T'$, each with a Stinespring dilation
  chosen as in eq.~(\ref{eq:embedding}), $C(T\otimes T')=C(T)+C(T')$, then
  $$\forall\rho\in{\cal O}_T,\rho'\in{\cal O}_{T'}\quad
                                E_f(\rho\otimes\rho')=E_f(\rho)+E_f(\rho').$$
  \qed
\end{thm}
\par
Most interesting is the case when we know $C(T^{\otimes n})=n C(T)$,
because then we can conclude $E_f(\rho^{\otimes n})=n E_f(\rho)$,
thus determining the entanglement cost of $\rho$ (see section~\ref{sec:entf}).
For example, King~\cite{king:unital:add,king:unital:add:2} proved this for
unital qubit--channels, Shor~\cite{Shor02} for entanglement--breaking
channels, and King~\cite{king:depol} for arbitrary depolarising channels,
giving rise to a host of states for which
we thus know that the entanglement cost equals $E_f$. Examples are discussed
in section~\ref{sec:examples} below and the following two sections.
\par\medskip
It is natural to consider ways to implement an implication of additivity
going the other way than theorem~\ref{thm:CC:implies:EE}: from entanglement of
formation to Holevo capacity.
\par
Indeed, in another look at eq.~(\ref{eq:hcap:alt}), let us focus on the
other quantity of interest
in the optimisation: this is the von Neumann entropy of the output state.
In general, while there can be many ensembles maximising eq.~(\ref{eq:hcap})
(let us assume for the moment that the output space is finite dimensional),
and in fact many averages $\sum_i p_i\pi_i$ (the set ${\cal O}_T$ of optimal
input states introduced above),
the average output state of such an optimal ensemble,
$\omega=\sum_i p_i T(\pi_i)$, is \emph{unique}: the reason is the strict concavity
of the von Neumann entropy, so if we had two ensembles with different
output states, mixing the ensembles would strictly increase the Holevo
mutual information. Let us denote this optimal output state $\omega(T)$.
\par
It is clear that the additivity conjecture eq.~(\ref{eq:hcap:add}) implies that 
\begin{equation}
  \label{eq:omega:mult}
  \omega(T\otimes T') = \omega(T)\otimes\omega(T'),
\end{equation}
but the reverse seems not obvious. Still, eq.~(\ref{eq:omega:mult}) might be a
reasonable first step towards proving additivity of $C(T)$ in general.
\par
Unfortunately, even assuming additivity of the entanglement of formation,
we have not been able to derive additivity of the channel
capacity from eq.~(\ref{eq:omega:mult}).
\par
However, let us assume that for the product channel $T\otimes T'$
an optimal \emph{input} state in eq.~(\ref{eq:hcap:alt})
is a product (due to the non--uniqueness of
optimal input states there might also be entangled ones!), $\rho\otimes\rho'$, say.
Then clearly, $E_f(\rho\otimes\rho') = E_f(\rho)+E_f(\rho')$ implies
$C(T\otimes T') = C(T)+C(T')$, in a reversal of the argument from
the proof of theorem~\ref{thm:CC:implies:EE}.

\section{Superadditivity: unifying $C(T)$ and $E_f(\rho)$}
\label{sec:superadd}
Looking at eq.~(\ref{eq:hcap:alt}), and trying to find a unifying
reason why both of the above discussed additivity conjectures should
hold, we are led to speculate that $E_f$ might not only be additive
with respect to tensor products (eq.~(\ref{eq:entf:add})), but have
even a superadditivity property for arbitrary states on a composition
of two bipartite systems:
\par
Let $\rho$ be a state on ${\cal H}\otimes{\cal H}'$, where
${\cal H}={\cal H}_1\otimes{\cal H}_2$ and
${\cal H}'={\cal H}_1'\otimes{\cal H}_2'$. Then superadditivity
means that
\begin{equation}
  \label{eq:entf:superadd}
  E_f(\rho) \geq E_f(\tr_{{\cal H}'}\rho)+E_f(\tr_{{\cal H}}\rho),
\end{equation}
where all entanglements of formation are understood with respect
to the $1$--$2$--partition of the respective system.
(This relation was apparently first considered in~\cite{VW},
and called \emph{strong superadditivity} there. We call it
just ``superadditivity'' here in simple analogy to, e.g.,
subadditivity of the von Neumann entropy.)
\par
Note that this implies additivity of $E_f$ when applied
to $\rho_1\otimes\rho_2$ since we remarked in section~\ref{sec:entf}
that the other inequality is trivial.
\par
Note on the other hand that it also implies additivity of $C(T)$,
by eq.~(\ref{eq:hcap:alt}) in section~\ref{sec:stine}:
by replacing a supposedly optimal $\rho$ on
${\cal K}\otimes{\cal K}'$ (for two channels $T$ and $T'$, and corresponding
Stinespring dilations which give rise to the subspaces ${\cal K}$ and ${\cal K}'$
in respective bipartite systems) by the tensor product of its marginals,
we can only increase the entropy (subadditivity), and only decrease the 
entanglement of formation (superadditivity).
\par
As an extension, let us show that it even implies an additivity formula
for the classical capacity under \emph{linear cost constraints}
(see~\cite{Holevo:cost}): in this problem, there is given a selfadjoint
operator $A$ on the input system, and a real number $\alpha$, additional
to the channel $T$. As signal states we allow only such states $\sigma$ on
${\cal H}^{\otimes n}$ for which $\tr(\sigma \widehat{A})\leq n\alpha+o(n)$,
with
$$\widehat{A}=\sum_{k=1}^n \1^{\otimes(k-1)}\otimes A\otimes \1^{\otimes(n-k)}.$$
(i.e., their average cost is asymptotically bounded by $\alpha$). Then it can be
shown~\cite{Holevo:cost,winter:qstrong} that the capacity $C(T;A,\alpha)$ in
the thus constrained system \emph{and using product states}
is given by a maximisation as in eq.~(\ref{eq:hcap}), only that the ensembles 
$\{p_i,\pi_i\}$ are restricted by $\sum_i p_i\tr(\pi_i A)\leq \alpha$.
(The same treatment applies if there are several linear cost inequalties of this kind. It
is only for simplicity of notation that we stick to the case of a single one.)
Because of the linearity of this condition in the states this yields a
formula for $C(T;A,\alpha)$ very similar to theorem~\ref{thm:hcap:entf}:
\begin{equation}
  \label{eq:hcap:cost:alt}
  C(T;A,\alpha) = \sup\{ S\bigl(\tr_{{\cal H}_1}\rho\bigr)-E_f(\rho) :
                                                     \rho\text{ state on }{\cal K},\ 
                                                     \tr(\rho A)\leq \alpha \}.
\end{equation}
By the general arguments given in previous sections we can conclude that this function
is concave in $\alpha$. The question of course is again, if entangled inputs
help to increase the capacity, or if
\begin{equation}
  \label{eq:hcap:cost:add}
  C\bigl(T^{\otimes n};\widehat{A},n\alpha) \stackrel{?}{=} nC(T;A,\alpha).
\end{equation}
We shall show that this indeed follows from the superadditivity, by showing
the following: for channels $T$, $T'$, cost operators $A$, $A'$, and cost
threshold $\widetilde{\alpha}$:
$$C\bigl(T\otimes T';A\otimes\1+\1\otimes A';\widetilde{\alpha}\bigr)
         =\sup_{\alpha+\alpha'=\widetilde{\alpha}}
            \bigl\{ C(T;A,\alpha)+C(T';A',\alpha') \bigr\}.$$
(Then, by induction and using the concavity, the equality in
eq.~(\ref{eq:hcap:cost:add}) follows.)
\par
Indeed, ``$\geq$'' is obvious by choosing, for $\alpha+\alpha'=\widetilde{\alpha}$,
optimal states $\rho$, $\rho'$ in the sense of eq.~(\ref{eq:hcap:cost:alt}),
and considering $\rho\otimes\rho'$. In the other direction, assume any optimal
$\omega$ for the product system, with marginal states $\rho$ and $\rho'$:
by definition,
$$\tr\bigl( (\rho\otimes\rho')(A\otimes\1+\1\otimes A') \bigr)
               =\tr\bigl( \omega(A\otimes\1+\1\otimes A') \bigr)
               \leq \widetilde{\alpha},$$
so also the product $\rho\otimes\rho'$ is admissible, and since there 
exist $\alpha$, $\alpha'$ summing to $\widetilde{\alpha}$ such that
$\tr(\rho A)\leq\alpha$, $\tr(\rho' A')\leq\alpha'$, the claim follows
in exactly the same way as for the unconstrained capacity.
\par
We have thus proved:
\begin{thm}
  \label{thm:superstrong}
  Superadditivity of $E_f$, eq.~(\ref{eq:entf:superadd}), implies
  additivity of entanglement of formation, of the Holevo
  capacity and of the Holevo capacity with cost constraint
  under tensor products.
  \qed
\end{thm}
\par\medskip
Observe the strong intuitive appeal of the superadditivity property:
it says that by measuring the entanglement via $E_f$, a system can only appear
less entangled if judged by looking at its subsystems individually.
Note that this is almost trivially true (by definition) for
the \emph{distillable entanglement}, while wrong for the
\emph{relative entropy of entanglement}~\cite{Vedral:Plenio}, because this
would make it an additive quantity, which we know it
isn't~\cite{VW,AEJPVM,AMVW}. The superadditivity also bears semblance to
a distributional property of the so--called \emph{tangle}~\cite{CKW}.
\par\medskip
Superadditivity is thus a very strong property.
If there is one ``nice'' underlying mathematical structure
to the additivity of $E_f$, it should indeed be this. Note that it is true if one of
the marginal states, say $\tr_{{\cal H}'}$, is separable: because then
its $E_f$ is $0$, and eq.~(\ref{eq:entf:superadd}) simply expresses the
monotonicity of $E_f$ under local operations (in this case: partial traces).
This was previously noted in~\cite{VW}.
\par
Observe that it is sufficient to prove superadditivity for a \emph{pure}
state $\rho=\ketbra{\psi}$, as then we can apply it to an optimal 
decomposition of $\rho$, together with the convex roof property,
eq.~(\ref{eq:entf:convroof}). This was apparently considered by
Benatti and Narnhofer~\cite{Benatti:Narnhofer:00}, who even conjectured
``good decompositions'' of the reduced states
$\tr_{{\cal H}}\ketbra{\psi}$ and $\tr_{{\cal H}'}\ketbra{\psi}$.
This latter conjecture however was refuted by
Vollbrecht and Werner~\cite{VW00} who constructed a counterexample.
\par
On the other hand, there is limited positive evidence in favour of superadditivity:
In~\cite{VDC}, eq.~(16), it is actually proved if the partial trace in one of the
subsystems is entanglement--breaking. We observed (following~\cite{VW})
that it is trivially true if one of the reduced states 
is separable. Some of our examples yield more cases of superadditivity.
E.g.~in example~\ref{expl:d-dim:depol} we constructed the subspaces
${\cal K}_\lambda$: for every pure state
$\psi\in{\cal K}_{\lambda_1}\otimes\cdots\otimes{\cal K}_{\lambda_n}$,
with reduced density operators $\rho_1,\ldots,\rho_n$
we get (using the additivity of the minimal output entropy
proved in~\cite{king:depol})
\begin{equation*}\begin{split}
  E(\psi) &\geq S_{\rm min}(T_1)+\ldots+S_{\rm min}(T_n) \\
          &=    E_f(\rho_1)+\ldots+E_f(\rho_n),
\end{split}\end{equation*}
the second line by the insight of example~\ref{expl:d-dim:depol} that
all states supported on ${\cal K}_{\lambda_i}$ have the same
entanglement of formation.
\par
Similarly, our other examples yield certain pure states for which we
obtain superadditivity.
\par
It seems to us that this question most elegantly sums up the two most
prominent additivity question in quantum information theory, and we would
like to pose it as a challenge:
either to prove superadditivity (thus proving additivity of $E_f$ and
of $C$), or to find a counterexample.

\section{Examples}
\label{sec:examples}
In this and the following two sections we want to demonstrate how
theorem~\ref{thm:CC:implies:EE} can be used to construct nontrivial
states for which we can compute the entanglement cost, to reproduce
some known results of this sort, and even exhibit ``irreversibility
of entanglement''.
\par
\begin{expl}
  \label{expl:general:depol}
  Consider the generalised depolarising channels of qubits:
  $$T:\rho\longmapsto\sum_{s=0,x,y,z} p_s\sigma_s\rho\sigma_s^\dagger,$$
  with $\sigma_0=\1$, the familiar Pauli matrices
  \begin{equation*}
    \sigma_x =\left(\begin{array}{rr}
                       0 & 1 \\
                       1 & 0
                    \end{array}\right),\ 
    \sigma_y =\left(\begin{array}{rr}
                      0 & -i \\
                      i &  0
                    \end{array}\right),\ 
    \sigma_z =\left(\begin{array}{rr}
                      1 &  0 \\
                      0 & -1
                    \end{array}\right),\ 
  \end{equation*}
  and a probability distribution $(p_s)_{s=0,x,y,z}$. For these channels additivity
  of the capacity under tensor product with an arbitrary channel
  was proved in~\cite{king:unital:add:2}.
  \par
  Note that up to unitary transformations on input and output system each unital
  qubit channel has this form, by the classification of qubit maps of King and
  Ruskai~\cite{king:ruskai}, and Fujiwara and Algoet~\cite{fujiwara}.
  By this result we also can assume that
  \begin{equation}
    \label{eq:king:ruskai}
    p_0+p_z-p_x-p_y \geq |p_0+p_y-p_x-p_z|,|p_0+p_x-p_y-p_z|.
  \end{equation}
  \par
  It is easy to see that for such a channel the capacity is given by
  $C(T)=1-S_{\rm min}(T)$, with the \emph{minimal output entropy}
  achieved at the eigenstates $\ket{0},\ket{1}$ of $\sigma_z$:
  $S_{\rm min}(T)=S\bigl(T(\ketbra{0})\bigr)=S\bigl(T(\ketbra{1})\bigr)$.
  An optimal ensemble is the uniform distribution on these states.
  \par
  It is easy to construct a Stinespring dilation for this map, by an isometry
  $U:\C^2\longrightarrow\C^2\otimes\C^4$, in block form:
  \begin{equation*}
    U=\left(\begin{array}{r}
              \sqrt{p_0}\sigma_0 \\
              \sqrt{p_x}\sigma_x \\
              \sqrt{p_y}\sigma_y \\
              \sqrt{p_z}\sigma_z
            \end{array}\right),
  \end{equation*}
  and the corresponding subspace ${\cal K}\subset\C^2\otimes\C^4$ is spanned by
  \begin{align*}
    \ket{\psi_T}       &=   \sqrt{p_0}\ket{0}\otimes\ket{0}
                          + \sqrt{p_x}\ket{1}\otimes\ket{x}
                          +i\sqrt{p_y}\ket{1}\otimes\ket{y}
                          + \sqrt{p_z}\ket{0}\otimes\ket{z}, \\
    \ket{\psi_T^\perp} &=   \sqrt{p_0}\ket{1}\otimes\ket{0}
                          + \sqrt{p_x}\ket{0}\otimes\ket{x}
                          -i\sqrt{p_y}\ket{0}\otimes\ket{y}
                          - \sqrt{p_z}\ket{1}\otimes\ket{z}.
  \end{align*}
  The optimal input state corresponds to the equal mixture $\rho_T$ of these
  two pure states.
  \par
  From these observations, together with theorem~\ref{thm:CC:implies:EE},
  we obtain that
  $$E_f(\rho_T)=S_{\rm min}(T)=H(p_0+p_z,1-p_0-p_z),$$
  and $E_f(\rho_T\otimes\sigma)=E_f(\rho_T)+E_f(\sigma)$ for any
  $\sigma\in{\cal O}_{T'}$, with arbitrary channel $T'$. In
  particular,
  $$E_c(\rho_T)=E_f(\rho_T)=H(p_0+p_z,1-p_0-p_z).$$
  \par
  In fact, we proved that the decomposition of $\rho_T^{\otimes n}$
  into the $2^n$ equally weighted tensor products of $\ketbra{\psi_T}$
  and $\ketbra{\psi_T^\perp}$ is formation--optimal. By the convex
  roof property of $E_f$ this implies that \emph{any} convex combination
  of these states is a formation--optimal decomposition (this argument was
  also used in~\cite{VW} to extend the domain of states with known entanglement
  of formation). In particular, we can conclude that \emph{any} mixture
  $\rho$ of $\ketbra{\psi_T}$ and $\ketbra{\psi_T^\perp}$ has
   \begin{equation}
     E_c(\rho)=E_f(\rho)=H(p_0+p_z,1-p_0-p_z).
   \label{eq:ec-qubit}
   \end{equation}
  \qed
\end{expl}
\par
The case of equal $p_x,p_y,p_z$ leads to the usual unitarily covariant
depolarising channel. This is contained in the following:
\begin{expl}
  \label{expl:d-dim:depol}
  Consider the $d$--dimensional depolarising channel with parameter $\lambda$:
  $$T:\rho\longmapsto \lambda\rho+(1-\lambda)\frac{1}{d}\1,$$
  with $-\frac{1}{d^2-1}\leq\lambda\leq 1$ for complete positivity, to ensure
  that $T$ can be represented as a mixture of generalised Pauli actions:
  $$T(\rho)=p_0\rho+(1-p_0)\sum_{i=1}^{d^2-1} \frac{1}{d^2-1} \sigma_i\rho\sigma_i^\dagger,$$
  with an orthogonal set of unitaries (a ``nice error basis'',
  see e.g.~\cite{werner:dense} for constructions) $\sigma_i$, i.e.
  $$\sigma_0=\1,\quad \tr(\sigma_i^\dagger\sigma_j)=d\delta_{ij},$$
  and $p_0=\lambda+(1-\lambda)/d^2$.
  \par
  For this channel, \cite{king:depol} proves the additivity of $C(T)$ and
  $S_{\rm min}(T)$, and it is quite obvious that
  $$C(T)=\log d - S_{\rm min}(T)=\log d - S\bigl(T(\ketbra{\psi})\bigr),$$
  for arbitrary $\ket{\psi}\in\C^d$, optimal input ensembles being those
  mixing to $\frac{1}{d}\1$. It is easy to evaluate this latter von Neumann
  entropy:
  \begin{equation*}\begin{split}
    S\bigl(T(\ketbra{\psi})\bigr)
    &=H\left(\lambda+\frac{1-\lambda}{d},\frac{1-\lambda}{d},\ldots,\frac{1-\lambda}{d}\right)\\
    &\!\!\!\!\!\!\!\!\!\!\!\!\!\!\!\!\!\!\!\!\!\!\!\!\!\!\!\!\!\!
     =H\left(\left(1-\frac{1}{d}\right)\!(1-\lambda),
              1-\left(1-\frac{1}{d}\right)\!(1-\lambda)\right)
        +\left(1-\frac{1}{d}\right)\!(1-\lambda)\log(d-1).
  \end{split}\end{equation*}
  \par
  Again, it is easy to construct a Stinespring dilation
  $U:\C^d\longrightarrow\C^d\otimes\C^{d^2}$ in block form:
  \begin{equation*}
    U=\left(\begin{array}{c}
              \sqrt{p_0}\1                           \\
              \sqrt{\frac{1-p_0}{d^2-1}}\sigma_1     \\
              \vdots                                 \\
              \sqrt{\frac{1-p_0}{d^2-1}}\sigma_{d^2-1}
            \end{array}\right),
  \end{equation*}
  such that the subspace of interest is $K_\lambda:=U\C^d$, its
  maximally mixed state denoted $\rho_\lambda$.
  Then theorem~\ref{thm:CC:implies:EE} allows us to conclude that
  $E_f(\rho_\lambda\otimes\sigma)=E_f(\rho_\lambda)+E_f(\sigma)$ for
  any $\sigma\in{\cal O}_{T'}$. In particular
  $$E_c(\rho_\lambda)=E_f(\rho_\lambda)=S_{\rm min}(T).$$
  By the argument familiar from example~\ref{expl:general:depol} we can conclude
  even that any mixture of product states on
  ${\cal K}_\lambda^{\otimes n}$ has entanglement of formation $n S_{\rm min}(T)$,
  in particular for every state $\rho$ supported on $K_\lambda$ we obtain
  $$E_c(\rho)=E_f(\rho)=S_{\rm min}(T).$$
  \qed
\end{expl}
\par\medskip
In the following section we will study some other examples, involving symmetry,
which allows evaluation of the entanglement of formation in some cases,
and also the entanglement cost.

\section{Group symmetry}
\label{sec:symmetry}
Imposing a group symmetry via representation on the involved
(sub--)spaces as follows, we obtain  another example, such as
 Vidal, D\"ur and Cirac~\cite{VDC},
and formulas valid for a class of channels.
Note that the symmetry is used 
principally for simplifying computations.
\par
Assume that a compact group $G$ (with Haar measure ${\rm d}g$) acts irreducibly
both on ${\cal K}$ and ${\cal H}_2$ by a unitary
representation (which we denote by $V_g$ and $U_g$), which commutes with
the map $T$ (partial trace):
\begin{equation}
  \label{eq:covariance}
  \tr_{{\cal H}_1}\bigl( V_g\sigma V_g^\dagger \bigr)
             = U_g\bigl( \tr_{{\cal H}_1}\sigma \bigr)U_g^\dagger.
\end{equation}
For example let there also be a unitary
representation of $G$ on ${\cal H}_1$, denoted $\widetilde{U}_g$,
such that ${\cal K}$ is an irreducible subspace of the
representation $V_g=\widetilde{U}_g\otimes U_g$.
We call this the \emph{Product Case}.
\par
In the general, non--product case of eq.~(\ref{eq:covariance}),
it is an easy exercise to show that, with $P$ denoting the
projection onto ${\cal K}$ in ${\cal H}_1\otimes{\cal H}_2$,
\begin{align}
  \label{eq:hcap:symmetry}
  C(T)  &= \log\dim{\cal H}_2-E_f\left(\frac{1}{\tr P}P\right), \\
  \label{eq:entf:symmetry}
  E_f\left(\frac{1}{\tr P}P\right)
        &= \min\bigl\{ E(\psi) : \ket{\psi}\in{\cal K} \bigr\}.
\end{align}
Indeed, in the second equation, ``$\geq$'' is trivially true, and for
the opposite direction choose a minimum entanglement pure state
$\ket{\psi_0}\in{\cal K}$, and consider the decomposition
$\{V_g\ketbra{\psi_0}V_g^\dagger,{\rm d}g\}$ of $\left(\frac{1}{\tr P}P\right)$
(by Schur's lemma!): all these states $V_g\ketbra{\psi_0}V_g^\dagger$
have the same entanglement,
\begin{equation}\begin{split}
\label{eq:smin-eq}
  E\bigl(V_g\ket{\psi_0}\bigr)
                 &= S\Bigl(\tr_1\bigl(V_g\ketbra{\psi_0}V_g^\dagger\bigr)\Bigr) \\
                 &= S\bigl(U_g\tr_1\ketbra{\psi_0}U_g^\dagger\bigr)             \\
                 &= S\bigl(\tr_1\ketbra{\psi_0}\bigr)=E(\psi_0),
\end{split}\end{equation}
using eq.~(\ref{eq:covariance}). As for the capacity, in the light of
eq.~(\ref{eq:hcap:alt}) and using eq.~(\ref{eq:entf:symmetry}),
the ``$\leq$'' is trivial, and the argument just given proves equality.
\par\medskip
Moreover, for \emph{all} states $\rho$ spanned by
$\{V_g \ketbra{\psi_0} V_g^* : g\in G\}$, 
where $\ket{\psi_0}$  is a pure state with
$E(\ket{\psi_0}) = \min\bigl\{ E(\ket{\psi}) : \ket{\psi}\in{\cal K} \bigr\}$,
we can  conclude that
\begin{align*}
E_f(\rho) = \min\bigl\{ E(\psi) : \ket{\psi}\in{\cal K} \bigr\}.
\end{align*}
\par
We even obtain the entanglement cost of all the $\rho$ 
spanned by $\{V_g \rho_0 V_g^* : g\in G\}$,
in the cases where we know that
$E_c(\frac{1}{\tr P}P)=E_f(\frac{1}{\tr P}P)$:
consider the chain of inequalities
\begin{align*}
 E_f\left(\left(\frac{P}{\tr P}\right)^{\otimes n}\right)
                        &\leq \int {\rm d}^n g
                            E_f\bigl( V_{g_1}\otimes\cdots\otimes V_{g_n}
                                       \rho^{\otimes n}
                                      V_{g_1}^\dagger\otimes\cdots\otimes V_{g_n}^\dagger
                               \bigr)                                                     \\
                        &\leq \int {\rm d}^n g \sum_{k_1}^n
                            E_f\bigl( V_{g_k} \rho V_{g_k}^\dagger \bigr)                 \\
                        &= n E_f\left( \frac{P}{\tr P} \right)
                            =E_f\left( \left(\frac{P}{\tr P}\right)^{\otimes n} \right).
\end{align*}
Here the first inequality is due to the convexity (see the definition)
of $E_f$, applied to the family $V_g\rho V_g^\dagger$ with Haar measure,
and the others are by subadditivity of $E_f$ and the assumption.
But the right hand side in the first line equals $E_f\bigl(\rho^{\otimes n}\bigr)$,
since any decomposition of $\rho^{\otimes n}$ translates into
a decomposition of $V_{g_1}\otimes\cdots\otimes V_{g_n}\rho^{\otimes n}
V_{g_1}^\dagger\otimes\cdots\otimes V_{g_n}^\dagger$ of the same entanglement,
and vice versa. Hence
\begin{align}
  E_c(\rho)=E_f(\rho)=\min\bigl\{ E(\psi) : \ket{\psi}\in{\cal K} \bigr\}.
  \label{eq:ef=ec=smin} 
\end{align}
\par
(Note that in~\cite{VDC} this was argued by making use of being
in the ``product case'', in which case the group action on
${\cal K}$ is performable by LOCC; then the first inequality
above was argued by nonincrease of $E_f$ under LOCC transformations.)

In particular, if in addition the action of $G$ in ${\cal K}$ is 
{\it transitive}, we can conclude
(\ref{eq:ef=ec=smin} ) for all the state supported on ${\cal K}$,
because (\ref{eq:smin-eq}) implies 
that  
$E(\ket{\psi})$ takes the same value
for any pure state $\ket{\psi}$ in ${\cal K}$.
\par
This group symmetry argument simplifies the analysis of 
unital qubit channels and generalised depolarising channels.
In the former case, $G$ is chosen to be $SU(d)$, 
while in the latter, we consider  the group $G=\{\1, R, R^2, R^3 \}$, with
\begin{align*}
 R=\left(\begin{array}{cc}
  0& -1 \\
  1 & 0
 \end{array}\right).
\end{align*}
In both cases, we define representations $V_g, U_g$ of $G$ 
by $V_g= U g U^*$ and $U_g=g$.
They are irreducible, and satisfy the condition eq.~(\ref{eq:covariance}).
Hence,  general arguments in this section,
directly implies results about these examples in the previous section.
\par
The following example is constructed using group symmetry.
\begin{expl}
  \label{expl:VDC}
  Vidal, D\"ur and Cirac~\cite{VDC} consider the subspace ${\cal K}$
  of $\C^3\otimes\C^6$ spanned by
  \begin{align*}
    \ket{0}_s &= \frac{1}{2}\bigl(\ket{1}\ket{2}+\ket{2}\ket{1}+\sqrt{2}\ket{0}\ket{3}\bigr), \\
    \ket{1}_s &= \frac{1}{2}\bigl(\ket{2}\ket{0}+\ket{0}\ket{2}+\sqrt{2}\ket{1}\ket{4}\bigr), \\
    \ket{2}_s &= \frac{1}{2}\bigl(\ket{0}\ket{1}+\ket{1}\ket{0}+\sqrt{2}\ket{2}\ket{5}\bigr).
  \end{align*}
  By using the isomorphism $\ket{j}\leftrightarrow\ket{j}_s$ between $\C^3$ and
  ${\cal K}$, it is easily checked that $\tr_{\C^6}$ implements the channel map
  $$T:\rho\longmapsto\frac{1}{4}\bigl(\1+\rho^\top\bigr),$$
  hence we are in the transitive covariant case, with $U\in{\rm SU}(3)$ and
  $V=\overline{U}$. It is straightforward to check that this channel
  is entanglement--breaking (see~\cite{VDC}): hence~\cite{Shor02}
  tells us that its capacity is additive, and we can apply
  theorem~\ref{thm:CC:implies:EE}.
  \par
  By our general observations above we can conclude that for any state $\rho$
  supported on ${\cal K}$, $E_c(\rho)=E_f(\rho)=3/2$.
  \par
  Following~\cite{VDC}, we can introduce (for $j=0,1,2$)
  $$\ket{j}_t=\ket{\Phi_3}\otimes\ket{j}\in\C^3\otimes\C^3\otimes\C^3,$$
  and form the superpositions
  $$\ket{\widetilde{\jmath}}:=c\ket{j}_s \oplus s\ket{j}_t
                                         \in\C^3\otimes\bigl(\C^6\oplus\C^9\bigr)$$
  in the direct sum of the respective supporting spaces, with $|c|^2+|s|^2=1$.
  This obviously retains the covariant nature, and allows us to implement
  the mixtures of $T$ with the constant map onto $\frac{1}{3}\1$, so we get
  every channel
  $$T_p:\rho\longmapsto p\frac{1}{3}\1+(1-p)\rho^\top,$$
  for $3/4\leq p\leq 1$, all of which are clearly entanglement--breaking,
  so the same technique applies, and we find subspaces on which every state
  has $E_c=E_f={\rm const.}\in[3/2,\log 3]$.
  \qed
\end{expl}
\par
In~\cite{VDC}, by implementing other entanglement--breaking
channels (and using Shor's result~\cite{Shor02} on capacity additivity),
other, and more general results of this type were obtained.
\par
\begin{expl}
  \label{expl:antisymmetric}
  The ``$U\otimes U$''--representation of ${\rm SU}(3)$ 
  on $\C^3\otimes\C^3$ decomposes
  into two irreducible parts, the symmetric subspace of dimension $6$ and the
  antisymmetric subspace ${\cal A}$ of dimension $3$.
  The latter has a nice basis given by
  \begin{align*}
    \ket{0}_a &=\frac{1}{\sqrt{2}}\bigl(\ket{1}\ket{2}-\ket{2}\ket{1}\bigr), \\
    \ket{1}_a &=\frac{1}{\sqrt{2}}\bigl(\ket{2}\ket{0}-\ket{0}\ket{2}\bigr), \\
    \ket{2}_a &=\frac{1}{\sqrt{2}}\bigl(\ket{0}\ket{1}-\ket{1}\ket{0}\bigr),
  \end{align*}
  which we use to identify ${\cal A}$ with $\C^3$.
  \par
  Notice that the partial trace over the first factor (say) implements a
  unital channel with symmetry ($U\in{\rm SU}(3)$ on $\C^3$ and $V=U\otimes U$
  on ${\cal A}$), which is even transitive (hence all states $\rho_a$
  supported on ${\cal A}$ have the same entanglement of formation
  $E_f(\rho_a)=1$), but it
  is neither depolarising nor entanglement--breaking:
  in the above identification it reads
  $$T_{\rm VDC}:\rho_a\longmapsto \frac{3}{2}\left(\frac{1}{3}\1\right)
                                               -\frac{1}{2}\rho^\top.$$
  Notice that this is one of the very channels used in~\cite{Werner:Holevo}
  to disprove the general multiplicativity conjecture for the
  \emph{maximal output $p$--norm} of a channel.
  Incidentally, this property is the main tool in King's proofs of the additivity
  of channel capacities~\cite{king:unital:add,king:unital:add:2,king:depol}.
  \qed
\end{expl}
\par
Denoting the maximally mixed state on ${\cal A}$ by $\sigma_{\cal A}$,
it was shown in~\cite{Shimono} that 
$E_f(\sigma_{\cal A}\otimes\sigma_{\cal A})=2E_f(\sigma_{\cal A})=2$.
Subsequently, Yura~\cite{yura} has shown that for all $n$,
$E_f\left( \sigma_{\cal A}^{\otimes n} \right)=n$, showing that the entanglement
cost of this state is indeed $1$.
\par\medskip
The above examples show that using covariance one can often evaluate the
entanglement of formation. By carefully choosing the supporting subspace of
the state we can use our main theorem~\ref{thm:CC:implies:EE}, yielding
even the entanglement cost.

\section{Gap between $E_c$ and $E_D$}
\label{sec:gap}
  Returning to example~\ref{expl:general:depol}, let us demonstrate that 
  the states discussed there exhibit a gap between the entanglement cost
  and distillable entanglement for some of these states,
  by use of the $\log$--negativity bound $\log\|\rho^{\Gamma}\|_{1}$
  on distillable entanglement~\cite{Vidal:Werner}.
  \par
  We use the notation of example~\ref{expl:general:depol}, in particular we
  assume the channel $T$ to be a mixture of Pauli rotations, with
  probability weights according to eq.~(\ref{eq:king:ruskai}).
  The partial transpose $\rho_T^{\Gamma}$ of the optimal state
  $\rho_T$ decomposes into a direct sum of two $4\times 4$--matrices,
  which turn out to have the same characteristic equation
  \begin{align*}
    f(2z) &= 0,\ \text{ where} \\
    f(z)  &= z^4 - z^3 + 4(p_0p_xp_y+p_0p_xp_z+p_0p_yp_z+p_xp_yp_z)z - 16 p_0p_xp_yp_z.
  \end{align*}
  Since $f\left(2z\right)=0$ has only one negative root $z_0$ and 
  $f$ is decreasing in a neighbourhood of it,
  $\log\|\rho_T^{\Gamma}\|_{1}< E_c(\rho_T)$ is equivalent to
  \begin{equation}
    f\left(-\frac{2^{E_c(\rho_T)}-1}{2}\right)
      =f\left(-\frac{2^{H(p_0+p_z,1-p_0-p_z)}-1}{2}\right) > 0,
    \label{eq:ec>ed}
  \end{equation}
  using $\left\|\rho_T^\Gamma\right\|_1=1-4z_0$.
  \par
\begin{figure}[ht]
    \centering
    \includegraphics{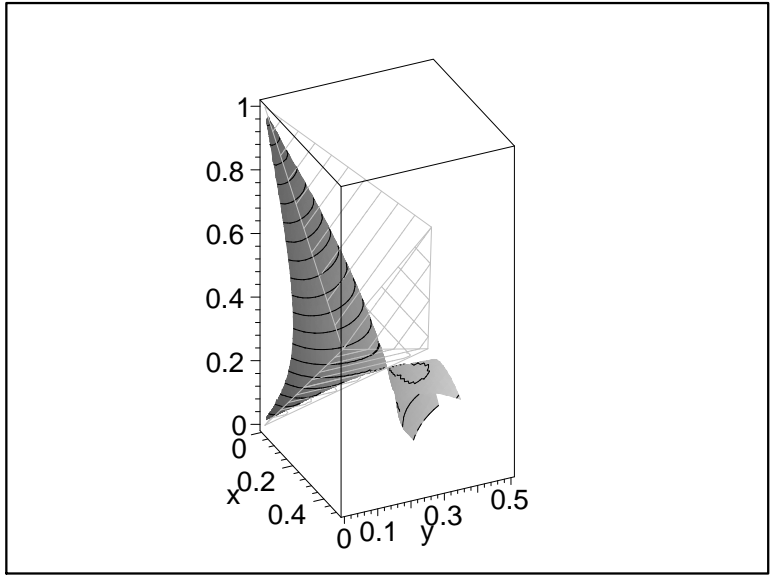}
    \caption{\small Plots in a $(p_x,p_y,p_z)$--frame of the admissible parameters
      according to eq.~(\ref{eq:king:ruskai}) and of the region for which
      eq.~(\ref{eq:ec>ed}) holds (between the two surfaces).}
    \label{fig:1}
  \end{figure}

  That is, if $p_0$, $p_x$, $p_y$, $p_z$ satisfy this inequality,
  there is a gap between the entanglement cost of $\rho_T$,
  and its distillable entanglement; figure~\ref{fig:1} shows a plot
  of the region of these $(p_x,p_y,p_z)$. By continuity, also for a mixture of 
  $\ketbra{\psi_T}$ and $\ketbra{\psi_T^{\perp}}$ which is sufficiently
  close to $\rho_T$, we observe a similar gap.
  \par
  Especially, for $p_0=1/2$, $p_x=p_y=p_z=1/6$,
  a short calculation reveals that $\left\|\rho_T^{\Gamma}\right\|_1=5/3$,
  so $E_D(\rho_T)\leq\log(5/3)\approx 0.737$, which is smaller
  than the entanglement cost $E_c(\rho_1)=H(1/3,2/3)\approx 0.918$.
  \par
  If $p_0+p_z=p_x+p_y=\frac{1}{2}$ and $p_0\neq p_z$, $p_x\neq p_y$,
  we can even prove for \emph{all} true mixtures 
  $\rho_{T,s}= s\ketbra{\psi_T}+(1-s)\ketbra{\psi_T}^\perp$  of 
  $\ketbra{\psi_T}$ and $\ketbra{\psi_T^{\perp}}$,
  that $E_D(\rho_{T,s})<E_c(\rho_{T,s})$ holds:
  by eq.~(\ref{eq:ec-qubit}) the latter is
  $1$ for all these $\rho_{T,s}$, and
  the key observation is that 
  $\log\|\rho_T^{\Gamma}\|_{1}$ is strictly smaller than 
  $E_c(\rho_{T,s})$ in this case,
  for the conditon~(\ref{eq:ec>ed}) is always satisfied. Hence
  $\left\| \rho_T^\Gamma \right\|_1 < 2$.
  \par
  The convexity of trace norm and the observation
  $\left\|\ketbra{\psi_T}^{\Gamma}\right\|_1=2$
  leads, for $\frac{1}{2}\leq s<1$ (which we may assume by symmetry), to
  \begin{equation*}\begin{split}
    \left\|\rho_{T,s}^{\Gamma}\right\|_1
       &\leq (2s-1)\left\|\ketbra{\psi_T}^\Gamma\right\|_1
            +(2-2s)\left\|\rho_{T}^{\Gamma}\right\|_1           \\
       &<    (2s-1)\cdot 2                   +(2-2s)\cdot 2 = 2,
  \end{split}\end{equation*}
  and consequently we have 
  $\log\left\|\rho_{T,s}^{\Gamma}\right\|_{1} < 1 = E_c(\rho_T)$.
  \par
  This fact is also proven by noting that the negativity is 
  \emph{strictly} convex for 
  mixings of $\ketbra{\psi_T}$ and $\ketbra{\psi_T^{\perp}}$, i.e.
  $$\left\| \bigl(s \ketbra{\psi_T}+(1-s)\ketbra{\psi_T^{\perp}}\bigr)^{\Gamma} \right\|_1
         < s \left\| \ketbra{\psi_T}^{\Gamma}   \right\|_1
          + (1-s)\left\|\ketbra{\psi_T^{\perp}}^{\Gamma}    \right\|_1,$$
  if $0<s<1$. 
  This is proved by finding eigenvectors with nonzero overlap
  of the two partial transposes such that one has a negative, the other
  a positive eigenvalue.

\section{Conclusion}
\label{sec:conclusion}
We demonstrated a link between the additivity problems for classical capacity of
quantum channels and entanglement of formation, resulting in the additivity
of the latter for many states, by invoking recent additivity results for the former.
This allows us to establish in particular a gap between distillable entanglement and
entanglement cost for many of these states. By exploiting the fact that $E_f$
is a convex roof, this additivity can be extended to even more states, though it
is not clear how far this would get us, even taking the general additivity conjecture
for granted.
\par
It is obvious that we only probed the scope of the method, and it is clear
that other examples of the same sort can be constructed, adding to the list of
states for which the entanglement cost is known. Each channel for which
additivity of its capacity is established will add to this list.
\par
The method generalises part of the argument found in the recent work of
Vidal, D\"ur and Cirac~\cite{VDC}, but for the case of entanglement breaking channels
their method is more general.
\par
On the side of general insights, the attempt to link the two additivity conjectures
considered here led us to consider the superadditivity of entanglement of formation
as a relation which integrates them neatly. We were even able to exhibit a few cases
where it is known to hold, providing modest evidence in favour of it.
\par\bigskip\noindent
{\bf Since completion of this work,}
subsequent research has further clarified the picture presented here:
Ruskai ({\tt quant-ph/0303141}) showed that not all bipartite states
can be associated with a channel such as to make use of
theorem~\ref{thm:CC:implies:EE} to prove additivity.
Audenaert and Braunstein ({\tt quant-ph/0303045}) have re--expressed
the superadditivity of entanglement of formation using tools from
convex analysis, and showed that the multiplicativity conjecture
for maximal output $p$--norms~\cite{AHW},
for $p$ close to $1$, of filtering operations implies superadditivity.
Shor ({\tt quant-ph/0305035}) has complemented our theorem~\ref{thm:superstrong}
by showing that the general conjectures of superadditivity of $E_f$,
additivity of $E_f$ under tensor products, and additivity of $C$
are in fact equivalent to each other and to the additivity of minimal output
entropy of a channel.

\section*{Acknowledgements}
We thank K. G. Vollbrecht and R. F. Werner for conversations about the superadditivity
conjecture, and A. Uhlmann for pointers to the literature.
\par
KM and TS are supported by the Japan Science and Technology Corporation,
AW is supported by the U.K. Engineering and Physical Sciences Research Council,
and gratefully acknowledges the hospitality of the ERATO Quantum Computation
and Information project, Tokyo, on the occasion of a visit during which part
of the present work was done.


\begin{thebibliography}{M} \small
  \bibitem{AHW} G. G. Amosov, A. S. Holevo, R. F. Werner, ``On the additivity hypothesis
    in quantum information theory'' (Russian), Problemy Peredachi Informatsii, vol. 36,
    no. 4, pp. 25--34, 2000. English translation in Probl. Inf. Transm., vol. 36, no. 4,
    pp. 305--313, 2000.

  \bibitem{AH} G. G. Amosov, A. S. Holevo, ``On the multiplicativity conjecture for
    quantum channels'', Theor. Probab. Appl., vol. 47, no. 1, pp. 143--146, 2002.

  \bibitem{AEJPVM} K. Audenaert, J. Eisert, E. Jane, M.B. Plenio, S. Virmani,
    B. De Moor, ``The asymptotic relative entropy of entanglement'',
    Phys. Rev. Letters, vol. 87, 217902, 2001.

  \bibitem{AMVW} K. Audenaert, B. De Moor, K. G. H. Vollbrecht, R. F. Werner,
    ``Asymptotic Relative Entropy of Entanglement for Orthogonally Invariant States'',
    Phys. Rev. A, vol. 66, 032310, 2002.

  \bibitem{Benatti:Narnhofer:00} F. Benatti, H. Narnhofer, ``On the Additivity
    of the Entanglement of Formation'', Phys. Rev. A, vol. 63, 042306, 2001.

  \bibitem{BNU} F. Benatti, H. Narnhofer, A. Uhlmann, ``Decompositions of Quantum States
    with Respect to Entropy'', Rep. Math. Phys., vol. 38, no. 1, pp. 123--141, 1996.

  \bibitem{BDSW} C. H. Bennett, D. P. DiVincenzo, J. A. Smolin, W. K. Wootters,
    ``Mixed--state entanglement and quantum error correction'', Phys. Rev. A, vol. 54,
    no. 5, pp. 3824--3851, 1996.

  \bibitem{BFS} C. H. Bennett, C. A. Fuchs, J. A. Smolin, ``Entanglement--Enhanced
    Classical Communication on a Noisy Quantum Channel'', in: Quantum Comminication, Computing,
    and Measurement (O. Hirota, A. S. Holevo, C. M. Caves eds.), pp. 79--88, Plenum,
    New York, 1997.

  \bibitem{bruss:etal} D. Bruss, L. Faoro, C. Macchiavello, M. Palma,
    ``Quantum entanglement and classical communication through a depolarising channel'',
    J. Mod. Optics, vol. 47, no. 2, pp. 325--331, 2000.

  \bibitem{CKW} V. Coffman, J. Kundu, W. K. Wootters, ``Distributed Entanglement'',
    Phys. Rev. A, vol. 61, 052306, 2000.

  \bibitem{CNT} A. Connes, H. Narnhofer, W. Thirring, ``Dynamical entropy of C${}^*$--algebras
    and von Neumann algebras'', Comm. Math. Phys., vol. 112, no. 4, pp. 691--719, 1987.

  \bibitem{fujiwara} A. Fujiwara, P. Algoet, ``One--to--one parametrization of
    quantum channels'', Phys. Rev. A, vol. 59, pp. 3290--3294, 1999.

  \bibitem{fujiwara:hashizume} A. Fujiwara, T. Hashizume, ``Additivity of the capacity
    of depolarizing channels'', Phys. Lett. A, vol. 299, no. 5/6, pp. 469--475, 2002.

  \bibitem{HHT} P. M. Hayden, M. Horodecki, B. M. Terhal, ``The asymptotic entanglement
    cost of preparing a quantum state'', J. Phys. A: Math. Gen., vol. 34, no. 35,
    pp. 6891--6898, 2001.

  \bibitem{Holevo73} A. S. Holevo, ``Some estimates for the amount of information
    transmittable by a quantum communications channel'' (Russian), 
    Problemy Peredachi Informatsii, vol. 9, no. 3, pp. 3--11, 1973.
    English translation: Probl. Inf. Transm., vol. 9, no. 3, pp. 177--183, 1973.

  \bibitem{Holevo77} A. S. Holevo, ``Problems in the mathematical theory of quantum
    communication channels'', Rep. Mathematical Phys., vol. 12, no. 2, pp. 273--278,
    1979.

  \bibitem{Holevo98} A. S. Holevo, ``The capacity of the quantum channel with
    general signal states'', IEEE Trans. Inf. Theory, vol. 44, no. 1,
    pp. 269--273, 1998.

  \bibitem{Holevo:cost} A. S. Holevo, ``On Quantum Communication Channels with
    Constrained Inputs'', e--print {\tt quant-ph/9705054}, 1997.

  \bibitem{king:unital:add} C. King, ``Maximization of capacity and $\ell_p$ norms
    for some product channels'', J. Math. Phys., vol. 43, no. 3,
    pp. 1247--1260, 2002.

  \bibitem{king:unital:add:2} C. King, ``Additivity for unital qubit
    channels'', J. Math. Phys., vol. 43, no. 10, pp. 4641--4653, 2002.

  \bibitem{king:depol} C. King, ``The capacity of the quantum depolarizing channel'',
    e--print {\tt quant-ph/0204172}, 2002.

  \bibitem{king:ruskai} C. King, M. B. Ruskai, ``Minimal Entropy of States Emerging
    from Noisy Quantum Channels'', IEEE Trans. Inf. Theory, vol. 47, pp. 192--209, 2001.

  \bibitem{kraus:book} K. Kraus, \emph{States, Effect and Operations: Fundamental Notions
     of Quantum Theory}, Springer Verlag, Berlin 1983.

  \bibitem{Rains} E. M. Rains, ``A Semidefinite Program for Distillable Entanglement'',
    IEEE Trans. Inf. Theory, vol. 47, no. 7, pp. 2921--2933, 2001.

  \bibitem{ruskai:review} M. B. Ruskai, ``Inequalities for Quantum Entropy: A Review with
    Conditions for Equality'', J. Math. Phys., vol. 43, pp. 4358--4375, 2002.

  \bibitem{SW:97} B. Schumacher, M. D. Westmoreland, ``Sending classical information
    via noisy quantum channels'', Phys. Rev. A, vol. 56, no. 1, pp. 131--138, 1997.

  \bibitem{Shimono} T. Shimono, ``Lower bound for entanglement cost of antisymmetric
    states'', e--print {\tt quant-ph/0203039}, 2002.

  \bibitem{evidence} P. W. Shor, J. A. Smolin, B. M. Terhal, ``Nonadditivity of
    Bipartite Distillable Entanglement follows from Conjecture on Bound Entangled
    Werner States'', Phys. Rev. Lett., vol. 86, pp. 2681--2684, 2001.

  \bibitem{Shor02} P. W. Shor, ``Additivity of the Classical Capacity of
    Entanglement--Breaking Quantum Channels'', e--print {\tt quant-ph/0201149}, 2002.

  \bibitem{Stinespring} W. F. Stinespring, ``Positive functions on $C^*$--algebras'',
    Proc. Amer. Math. Soc., vol. 6, pp. 211--216, 1955.

  \bibitem{TV} B. M. Terhal, K. G. H. Vollbrecht, ``The Entanglement of Formation
    for Isotropic States'', Phys. Rev. Letters, vol. 85, pp. 2625--2628, 2000.

  \bibitem{Vedral:Plenio} V. Vedral, M. B. Plenio, ``Entanglement measures and
    purification procedures'', Phys. Rev. A, vol. 57, no. 3, pp. 1619--1633, 1998.

  \bibitem{VDC} G. Vidal, W. D\"ur, J. I. Cirac, ``Entanglement cost of mixed
    states'', Phys. Rev. Letters, vol. 89, no. 2, 027901, 2002.

  \bibitem{Vidal:Werner} G. Vidal, R. F. Werner, ``A computable measure of entanglement'',
    Phys. Rev. A, vol. 65, 032314, 2002.

  \bibitem{VW00} K. G. H. Vollbrecht, R. F. Werner, ``A counterexample to a
    conjectured entanglement inequality'', e--print {\tt quant-ph/0006046}, 2000.

  \bibitem{VW} K. G. H. Vollbrecht, R. F. Werner, ``Entanglement Measures under
    Symmetry'', Phys. Rev. A, vol. 64, 062307, 2001.

  \bibitem{werner:dense} R. F. Werner, ``All teleportation and dense coding schemes'',
    J. Phys. A, vol. 34, no. 35, pp. 7081--7094, 2001.

  \bibitem{Werner:Holevo} R. F. Werner, A. S. Holevo, ``Counterexample to an additivity
    conjecture for output purity of quantum channels'', J. Math. Phys., vol. 43, no. 9,
    pp. 4353--4357, 2002.

  \bibitem{winter:qstrong} A. Winter, ``Coding Theorem and Strong Converse for
    Quantum Channels'', IEEE Trans. Inf. Theory, vol. 45, no. 7, pp. 2481--2485, 1999.

  \bibitem{winter:gen:add} A. Winter, ``Scalable programmable quantum gates and a new
    aspect of the additivity problem for the classical capacity of quantum channels'',
    e--print {\tt quant-ph/0108066}, 2001.

  \bibitem{wootters} W. K. Wootters, ``Entanglement of Formation of an Arbitrary State
    of Two Qubits'', Phys. Rev. Letters, vol. 80, no. 10, pp. 2245--2248, 1998.

  \bibitem{yura} F. Yura, ``Entanglement cost of three--level antisymmetric states'',
    J. Phys. A: Math. Gen., vol. 36, no. 15, pp. L237--L242, 2003.

\end{thebibliography}
\end{document}